\documentclass[twocolumn,showpacs,amsmath,amssymb,pr,superscriptaddress]{revtex4-2}
\usepackage{graphicx}
\usepackage{epsf}
\usepackage{ulem}
\usepackage{color}
\usepackage[unicode=true,colorlinks=true,urlcolor=blue,citecolor=blue]{hyperref}
\usepackage{xcolor}

\begin{document}



\title{AC Conductance in n-InSb Structures with Quantum Well. Acoustic Studies.} 

\author{I. L. Drichko} 
\affiliation{Ioffe Institute, Russian Academy of Sciences, St. Petersburg, 194021 Russia}
\author{I. Yu. Smirnov} 
\affiliation{Ioffe Institute, Russian Academy of Sciences, St. Petersburg, 194021 Russia}
\author{Yu. M. Galperin} 
\affiliation{Physics Institute, University of Oslo, Oslo, 0316 Norway}
\author{A. V. Suslov} 
\affiliation{National High Magnetic Field Laboratory, Tallahassee, FL 32310, USA}

\begin{abstract}
We studied the ac conductance of an $n$-InSb quantum well structure using acoustic methods in magnetic fields up to 18 T and at temperatures ranging from 20 to 500 mK. We attribute the unusual magnetic field dependences of surface acoustic wave (SAW) attenuation and velocity observed in the experiment to the presence of a conducting layer parallel to the quantum well in the sample. We successfully separated the contributions from both the quantum well and the shunting layer, enabling the identification of their distinct conduction mechanisms. Furthermore, by employing the coincidence technique, we determined the electron g-factor in the quantum well and investigated its dependence on the degree of spin polarization.
\end{abstract}

\maketitle

\section{Introduction} \label{introduction}

Indium antimonide is a unique platform for studying quantum effects due to its electrons having a small effective mass and a large g-factor value. This has enabled successful investigations of spin-orbit interaction in structures with InSb quantum wells~\citep{Zhi2025,Khodaparast2004}. However, the majority of research on such structures has focused on determining the g-factor~\citep{Chokomakoua2004,Nedniyom2009,Lei2020,Yang2011}.

Despite significant advances in the synthesis of modern semiconductor materials and low-dimensional systems, the quality of structures with n-InSb quantum wells still remains rather modest. While they exhibit relatively high mobility ($\sim 10^5$~cm$^2$/$(\mathrm{V} \cdot \mathrm{s})$) at temperatures $T <$4.2~K and demonstrate the integer quantum Hall effect, the anticipated transition to fractional quantum Hall states at lower temperatures remains elusive. To the best of our knowledge, no researchers have yet succeeded in observing such a crossover, even at $T <$0.1~K. Primarily, direct current methods, such as magnetotransport measurements, have been used to study these structures; however, acoustic methods have not been employed.

A review of the samples used in experiments revealed that the structure parameters typically fell within the range of carrier densities $n = (1\div3) \times 10^{11}$~cm$^{-2}$ and mobilities of (0.6$\div$2)$\times 10^5$~cm$^2$/$(\mathrm{V} \cdot \mathrm{s})$. Only in one study~\citep{Lehner2018} did the electron mobility reach a value of 3.5$\times 10^5$ cm$^2$/(V$\cdot$s). The sample we investigated had parameters within these stated limits.

With this work, we aimed to draw attention to the possible existence of a parallel conducting layer in such samples, which we have proven using an acoustic technique.

We have also shown that the presence of this conducting layer does not influence the determination of the electron g-factor in the quantum well by the coincidence method.

\section{Samples and methods}

Samples with a single $n$-InSb quantum well were grown using molecular beam epitaxy (MBE) on GaAs substrates. The structure consisted of 28 layers in total. In addition to this single QW, it contained buffers, doping, a cap, and other layers, designed to enhance the quality of the 2D electron gas.
The quantum well, with a width of 21~nm, is surrounded by  $\mathrm{In}_{0.9}\mathrm{Al}_{0.1}\mathrm{Sb}$ barriers and is located 150~nm below the top of the sample.

The $n$-type carriers were introduced to the active
region from the Si $\delta$-doping layer incorporated above the quantum well in the barrier (single-side doping).
The electron concentration in the well was $2.6 \times 10^{11}~\mathrm{cm}^{-2}$, and the mobility was $1.9 \times 10^{5}~\mathrm{cm}^2/(\mathrm{V} \cdot \mathrm{s})$ at $T=4.2~\mathrm{K}$.

\begin{figure}[h]
\centering
\includegraphics[width=0.9\linewidth]{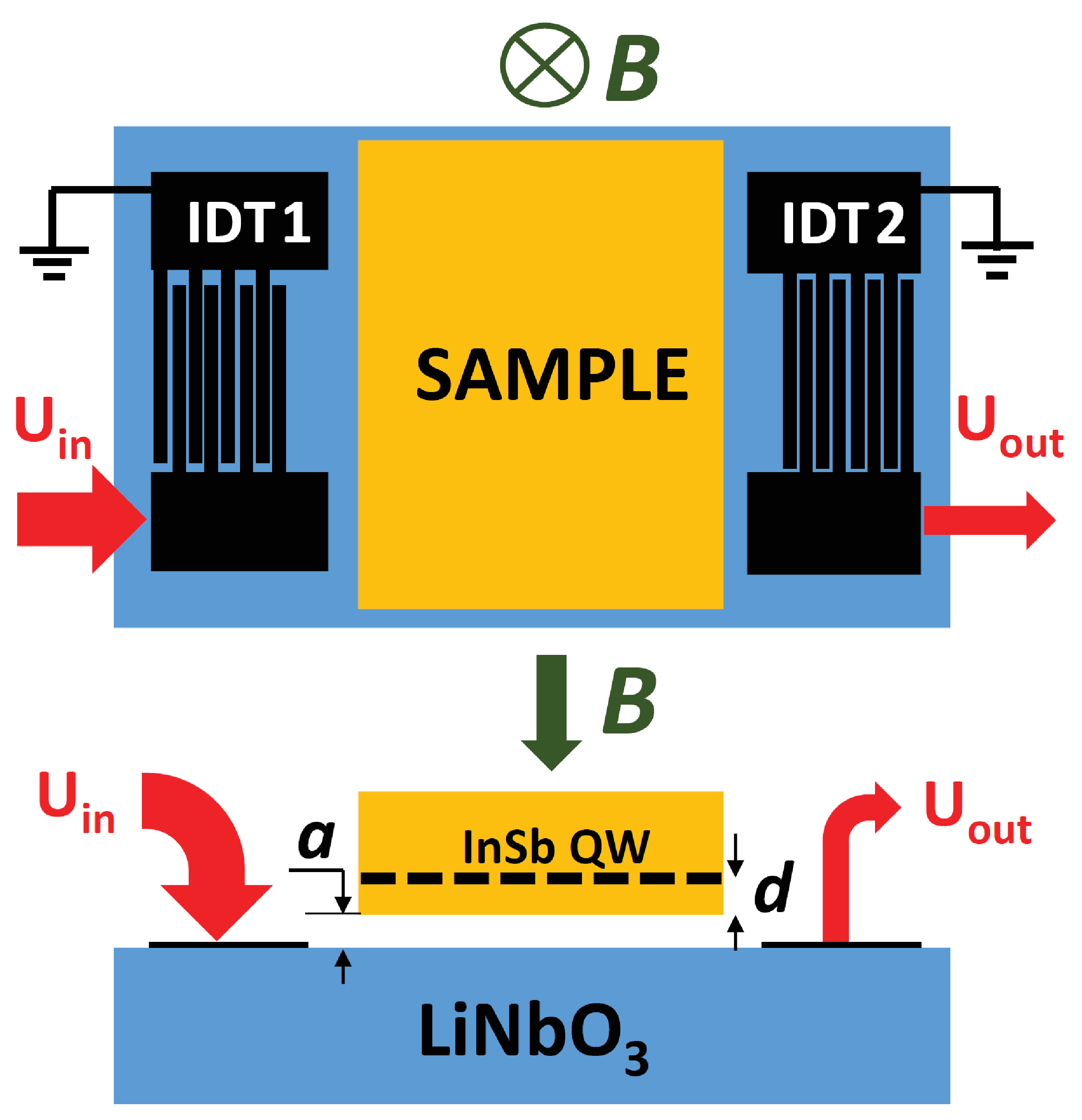}
\caption{Schematics of the experimental acoustic set-up: upper panel - top view; lower panel - side view.
}
\label{fig1}
\end{figure}

In Fig.~\ref{fig1} we show the schematic of the acoustic experimental setup. The sample is pressed by a spring to the surface of a lithium niobate (LiNbO$_3$) plate, on both sides of which interdigital transducers (IDTs) are patterned for surface acoustic wave (SAW) generation and detection.
A Y-cut lithium niobate platelet was used.
When an alternating voltage
$U_{\mathrm{in}}$   is applied to the input interdigital transducer (IDT1), a surface acoustic wave (SAW) is excited on the LiNbO$_3$ substrate due to the piezoelectric coupling; this wave propagated in the XZ-plane along the Z-direction. The SAW is accompanied by a traveling wave of ac electric field, also propagating along the Z-axis, which penetrates the two-dimensional channel, inducing alternating electric currents. These currents, in turn, give rise to Joule losses.

The interaction between the SAW's electric field and the electrons in the quantum well changes the amplitude and phase  of the SAW, which are detected by the second interdigital transducer (IDT2). Crucially,
in this experimental configuration, the mechanical deformation associated with the SAW is not transmitted to the studied system due to an effective clearance $a$
between the sample and LiNbO$_3$ substrate.

The acoustic setup was placed in a cryostat with a superconducting magnet. Measurements in a magnetic field of up to 18~T perpendicular to
the sample plane were carried out in a dilution refrigerator. We have measured the SAW attenuation ($\Gamma$) and variation of its velocity ($\Delta v/v$) at SAW frequencies of (MHz): 29, 85, 140, 252 and 306 in the transverse magnetic field range 0$<B<18$~T and temperature range of 20$\div$500 mK in a linear regime in the wave power.

The complex ac conductance, $\sigma^{\mathrm{ac}} (\omega) \equiv
\sigma_1 (\omega)-i\sigma_2 (\omega)$, is determined from simultaneous measurements of two values: the SAW attenuation, $\Gamma$, and the relative change in its velocity, $\Delta v/v$. The quantities $\Gamma$ and $\Delta v/v$ are related to the conductivity components through the following expressions~\citep{Kagan1997}:
\begin{eqnarray}
  \label{eq:G}
&&\Gamma=
8.68\frac{K^2}{2}qA     \frac{4\pi\sigma_1t(q)/\varepsilon_sv}
  {[1+4\pi\sigma_2t(q)/\varepsilon_sv]^2+[4\pi\sigma_1t(q)/\varepsilon_sv]^2}, \frac{\mathrm{dB}}{\mathrm{cm}}  \,   \\
&&\mathrm{where} \quad  A = 8b(q)(\varepsilon_1 +\varepsilon_0)
\varepsilon_0^2 \varepsilon_s
\exp [-2q(a+d)],\quad \mathrm{and}  \,   \nonumber \\
\label{eq:V}
&&\frac{\Delta v}{v}=\frac{K^2}{2}A    \frac{1+4\pi\sigma_2t(q)/\varepsilon_sv}
  {[1+4\pi\sigma_2t(q)/\varepsilon_sv]^2+[4\pi\sigma_1t(q)/\varepsilon_sv]^2},
\end{eqnarray}
where $K^2$ is the dimensionless piezoelectric coupling constant of  LiNbO$_3$, $q$ is the SAW's wave vector in lithium niobate, $d=$~150~nm is the depth at which the two-dimensional layer is located, $\varepsilon_1$, $\varepsilon_0$ and
$\varepsilon_s$ are the
dielectric constants of lithium niobate, vacuum,
and InSb respectively, $b$ and $t$ are complex functions of $a$, $d$, $\varepsilon_1$, $\varepsilon_0$ and
$\varepsilon_s$.
$a$ is the distance between the LiNbO$_3$ and the structure under study, determined from saturation of the SAW velocity shift in a strong magnetic field at $T$=20 mK~\citep{Drichko2011}.

\section{Experimental results in a perpendicular magnetic field and their interpretation}

\begin{figure}[t]
\centering
\includegraphics[width=0.9\linewidth]{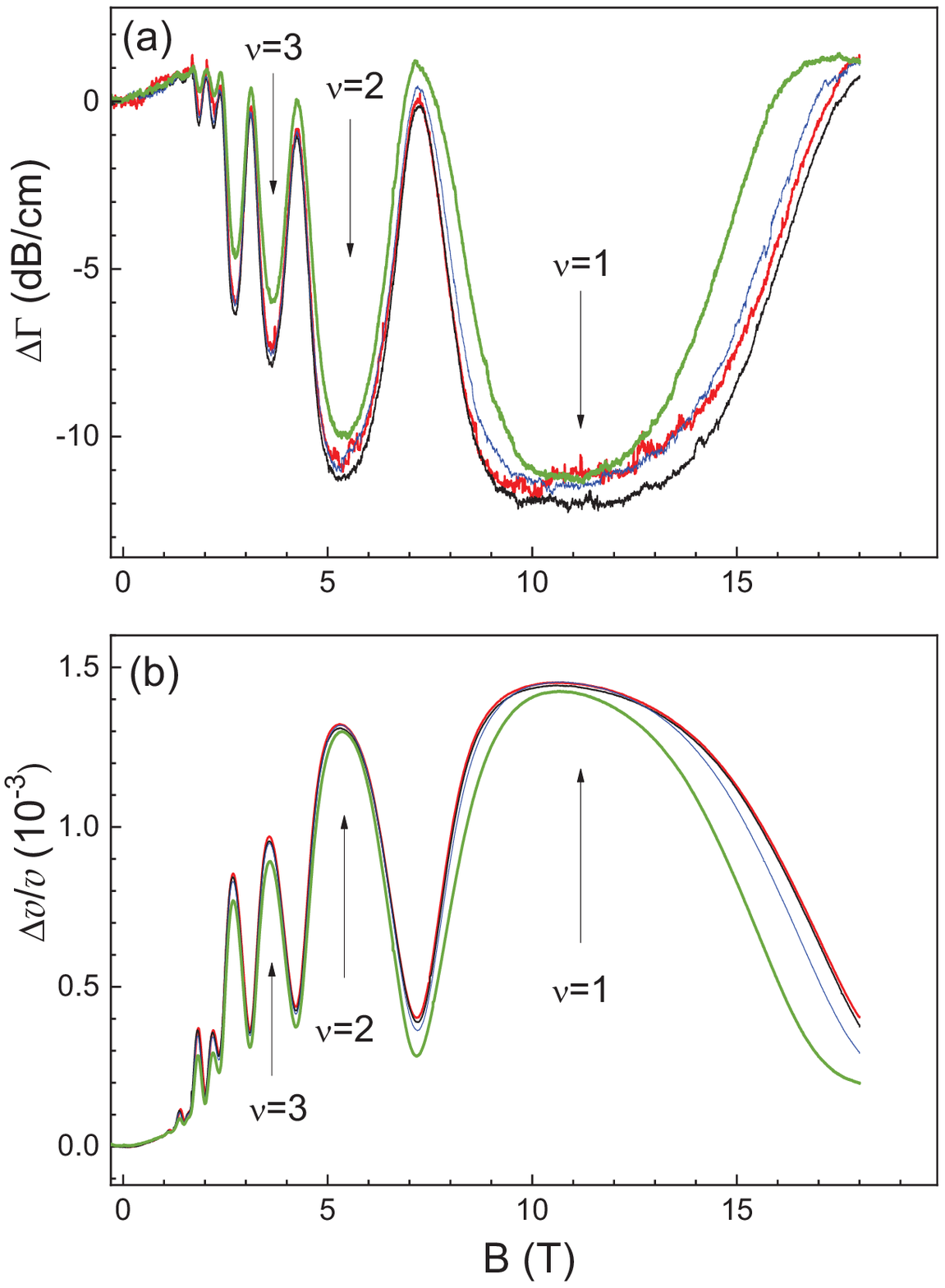}
\caption{Magnetic field dependences of $\Delta\Gamma = \Gamma(B) - \Gamma(B=0)$ (a) and $\Delta v/v$ (b) at different temperatures. $f$=85~MHz, $T$ (mK): 20 (red curve), 40 (black curve), 200 (blue curve), 500 (green curve).
Marked by arrows are the filling factors $\nu$.}
\label{fig2}
\end{figure}

Shown in Fig.~\ref{fig2}  are magnetic field dependences of $\Delta \Gamma$ and $\Delta v/ v$ at different temperatures.

The unusual aspect of our results is the \textit{negative} sign of the measured SAW attenuation change, $\Delta\Gamma(B)$, although the positions of the oscillation minima in the magnetic field correspond to the carrier density measured by the Hall effect. In high-quality quantum well samples, in accordance with theory~\citep{Kagan1997}, which has been repeatedly confirmed by experiment~\citep{Drichko2011,Drichko2000,Drichko2005,Drichko2013}, $\Delta\Gamma(B)$ always has a positive sign, and the magnitudes of the oscillation minima $\Delta\Gamma$ at even filling factors $\nu=2, 4, 6$ are close to 0. This fact leads to the conclusion that in the sample under study, the SAW attenuation by the electrons in the quantum well is observed on a specific negative background, whose magnitude depends on the magnetic field.

\begin{figure}[t]
\centering
\includegraphics[width=0.9\linewidth]{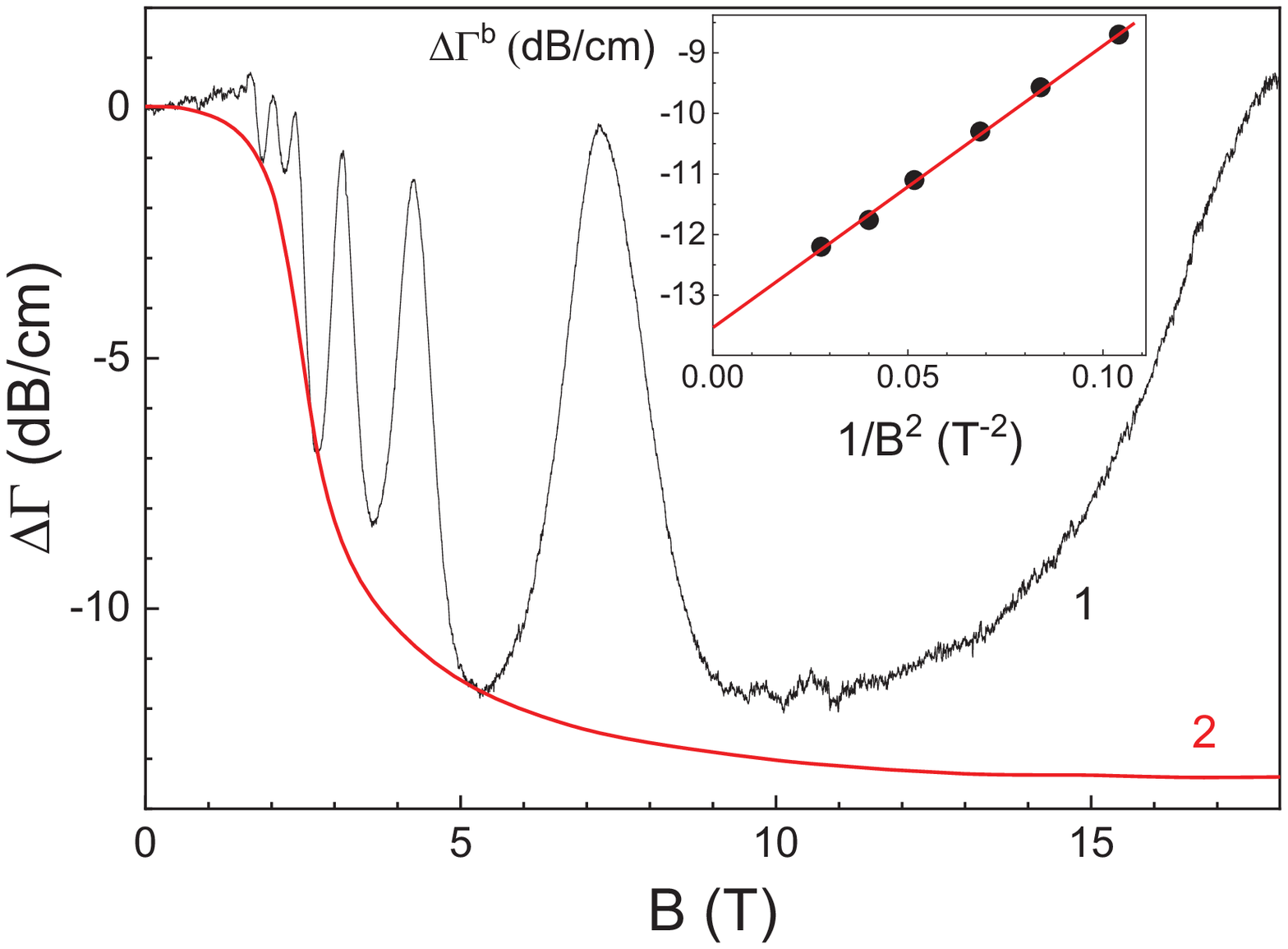}
\caption{1 - the experimental curve $\Delta \Gamma (B)$; 2 - the envelope curve $\Delta \Gamma^\text{b} (B)$. Inset: $\Delta \Gamma^\text{b}$ as  a function of $1/B^2$ in the region 3$\div$6~T. $T$=20~mK. $f$=85~MHz.
 }
\label{fig3}
\end{figure}

\begin{figure}[t]
\centering
\includegraphics[width=0.9\linewidth]{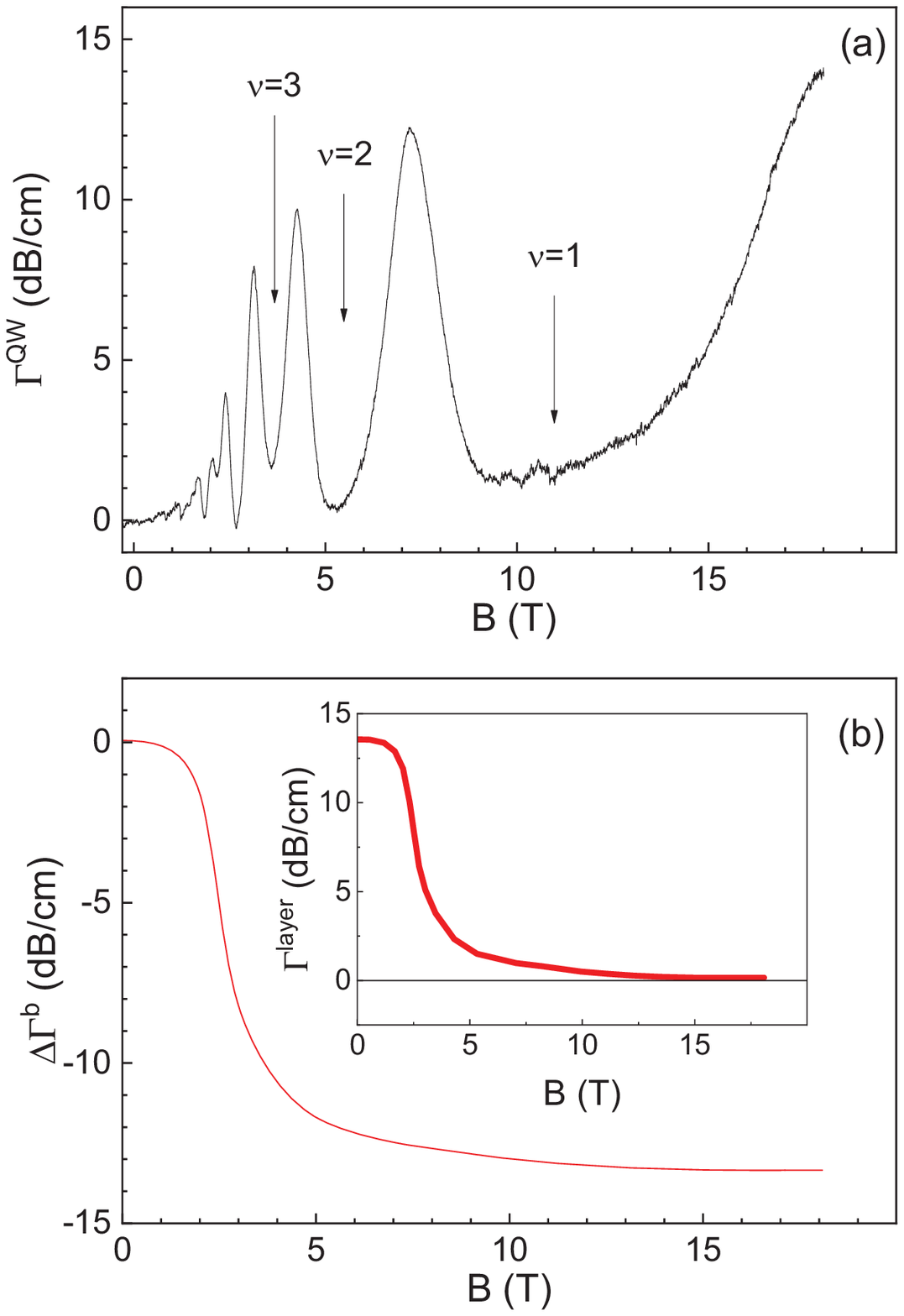}
\caption{a) The results of the subtraction: SAW absorption $\Gamma^{\text{QW}}(B)$ in the quantum well. b) Background SAW absorption $\Delta\Gamma^{b}(B)$. Inset: Absolute value of the SAW absorption in the shunting layer. $f$=85 MHz, $T$=20 mK.
}
\label{fig4}
\end{figure}

To determine the magnitude and magnetic field dependence of this background absorption $\Delta\Gamma^{\text{b}}(B)$, we draw an envelope curve through the experimental data points at magnetic fields $B=0$ and those corresponding to even filling factors $\nu=2, 4, 6$ (Fig.~\ref{fig3}). Then, to find the magnitude and magnetic field dependence of the SAW absorption by the electrons in the quantum well, it is necessary to subtract the background magnitude $\Delta\Gamma^{\text{b}}(B)$ from the experimental dependence $\Delta\Gamma(B)$. The result of this subtraction is presented in Fig.~\ref{fig4}.

Our acoustic measurements in a magnetic field are relative: in an experiment we measure the value $\Delta\Gamma=\Gamma(B)-\Gamma(B=0)$.  However, in our quantum well, which has a carrier mobility of ~10$^5$~cm$^2$/Vs, the zero-field conductivity is on the order of 10$^{-3}$~Ohm$^{-1}$. This corresponds to a negligible value of $\Gamma (B=0) \approx$~0. Consequently, the data presented in Fig.~\ref{fig4}a can be interpreted as the absolute attenuation $\Gamma^{\text{QW}}(B)$~\citep{Drichko2000,Drichko2005,Drichko2013}.

Since the sample is a multilayer structure, it is possible that the source of the background absorption is the interaction of the SAW with charge carriers with in one or several layers parallel to the quantum well. These layers have a low carrier density and mobility, with carriers being localized at low temperatures on impurity ions or in the minima of the random potential. In such layers, hopping ac conduction is realized.
As follows from both theory~\citep{Galperin1986} and experiment~\citep{Drichko2005a}, in the case of carrier hopping between localized states, $\Gamma(B)$  decreases with magnetic field, meaning that $\Delta\Gamma(B)$ has a negative sign. This behavior results from the magnetic freeze-out of carriers into localized states and the shrinkage of electron wave functions.

Furthermore, an analysis of the magnetic field dependence of this background absorption in the range of 3$\div$6~T, presented in the inset to Fig.~\ref{fig3}, shows that $-\Delta\Gamma^{\text{b}}(B) \propto 1/B^2$, which also agrees with the theory of SAW absorption for hopping ac conduction~\citep{Galperin1986}. Such a dependence  of $-\Delta\Gamma^{\text{b}}(B)$ on $1/B^2$ makes it possible, firstly, to determine the magnitude of the envelope curve at magnetic fields above 6~T (for $\nu < 2$) and, secondly, to determine the value of $\Gamma^{\text{b}}$ for $B=0$. Indeed, since $\Delta\Gamma^{\text{b}} = \Gamma(B) - \Gamma(B=0)$, and according to theory~\citep{Galperin1986} for hopping ac conduction $\Gamma(B) \rightarrow 0$ as $1/B^2 \rightarrow 0$, the intercept of the -$\Delta\Gamma^{\text{b}}(1/B^2)$ dependence on the Y-axis yields -$\Delta\Gamma^{\text{b}} = -\Gamma^{\text{b}}(B=0) = -13.5$ dB/cm  (see insert to Fig.~\ref{fig3}). Therefore, the absolute value of the  absorption $\Gamma^{\text{layer}}$ in the layer parallel to the quantum well is as shown in the inset to Fig.~\ref{fig4}b.

\begin{figure}[h]
\centering
\includegraphics[width=1\linewidth]{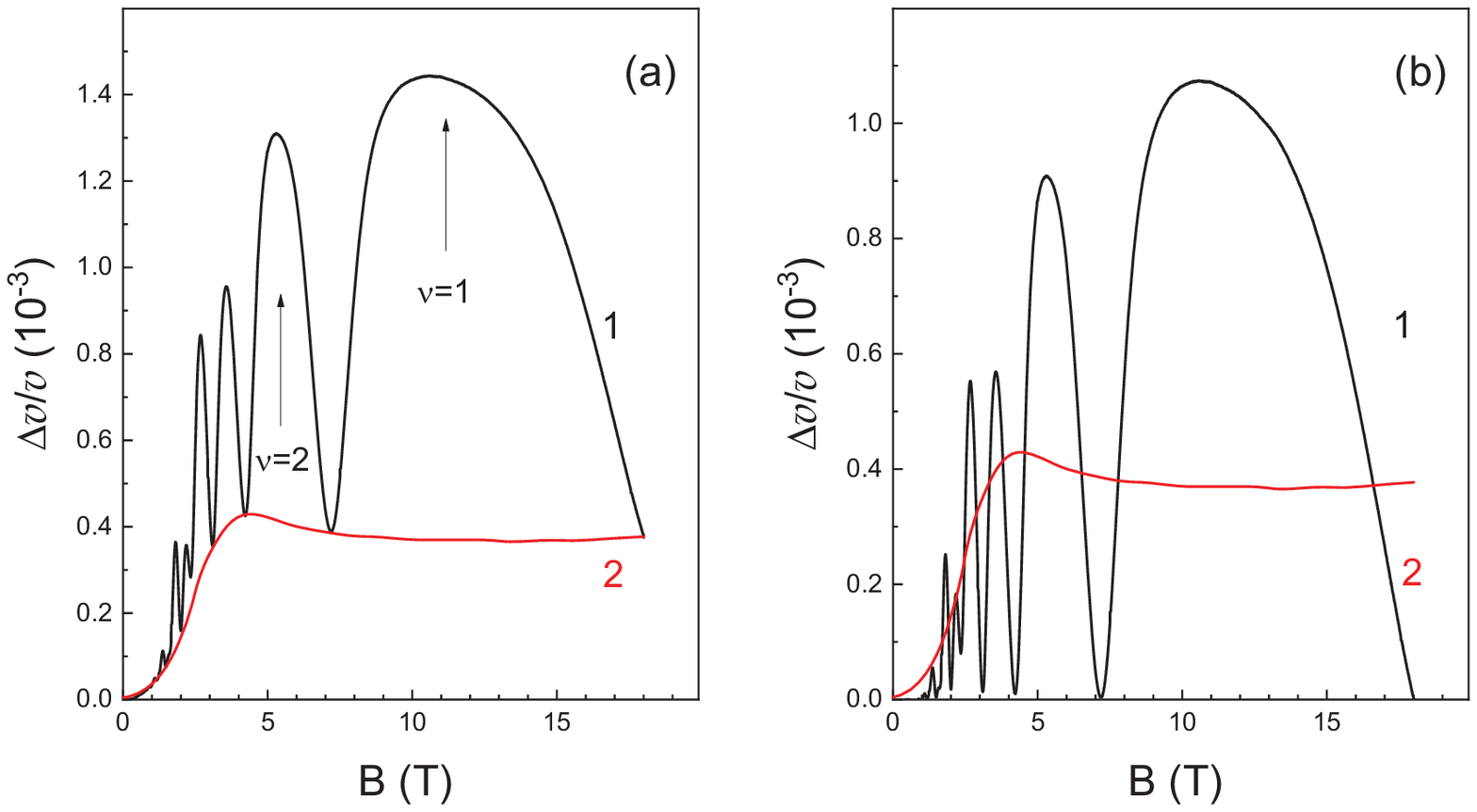}
\caption{Magnetic field dependence of the velocity change: a) 1 - experimental curve, black; 2 - envelope curve, red. b) 1 - After subtraction: $\Delta v/v$ contribution from the quantum well, black; 2 - $\Delta v/v$ contribution from the shunting layer, red; $T$=20~mK, $f$=85~MHz}
\label{fig5}
\end{figure}

To calculate the ac conductance components $\sigma_1$ and $\sigma_2$ in both the quantum well and the parallel layer  using Eqs.\ref{eq:G} and~\ref{eq:V}, in addition to $\Gamma (B)$, the magnetic field dependences of the SAW velocity change ($\Delta v/v$) must also be known for each of them separately. Figure~\ref{fig5}a presents the experimentally observed dependence $\Delta v/v (B)$.
Based on numerous experiments~\citep{Drichko2011,Drichko2000,Drichko2005,Drichko2013}, we assume that, similar to the attenuation,  the $\Delta v/v$ in the quantum well is also superimposed on a background originating from the parallel layer.

To obtain this background, the envelope curve in this case (see Fig.~\ref{fig5})  was constructed based on  previous experimental data:
in the case of SAW interaction with carriers in the quantum well the value of $\Delta v/v$ at its minima approaches~0~\citep{Drichko2011,Drichko2000,Drichko2005,Drichko2013}. Note that the characteristic shape of the $\Delta v/v$ envelope (curve 2) versus magnetic field is qualitatively similar to $\Delta v/v (B)$ in the case of hopping conductance measured in ~\citep{Drichko2005a}.
After subtracting the background from the experimental curve, we obtain the $\Delta v/v$ for the quantum well (1) and the shunting layer (2), as shown in Fig.~\ref{fig5}(b).

Additionally, it should be noted that the conductances in both the quantum well and the shunting layer show almost no temperature dependence up to 500~mK (Fig.~\ref{fig2}).

\begin{figure}[h]
\centering
\includegraphics[width=0.9\linewidth]{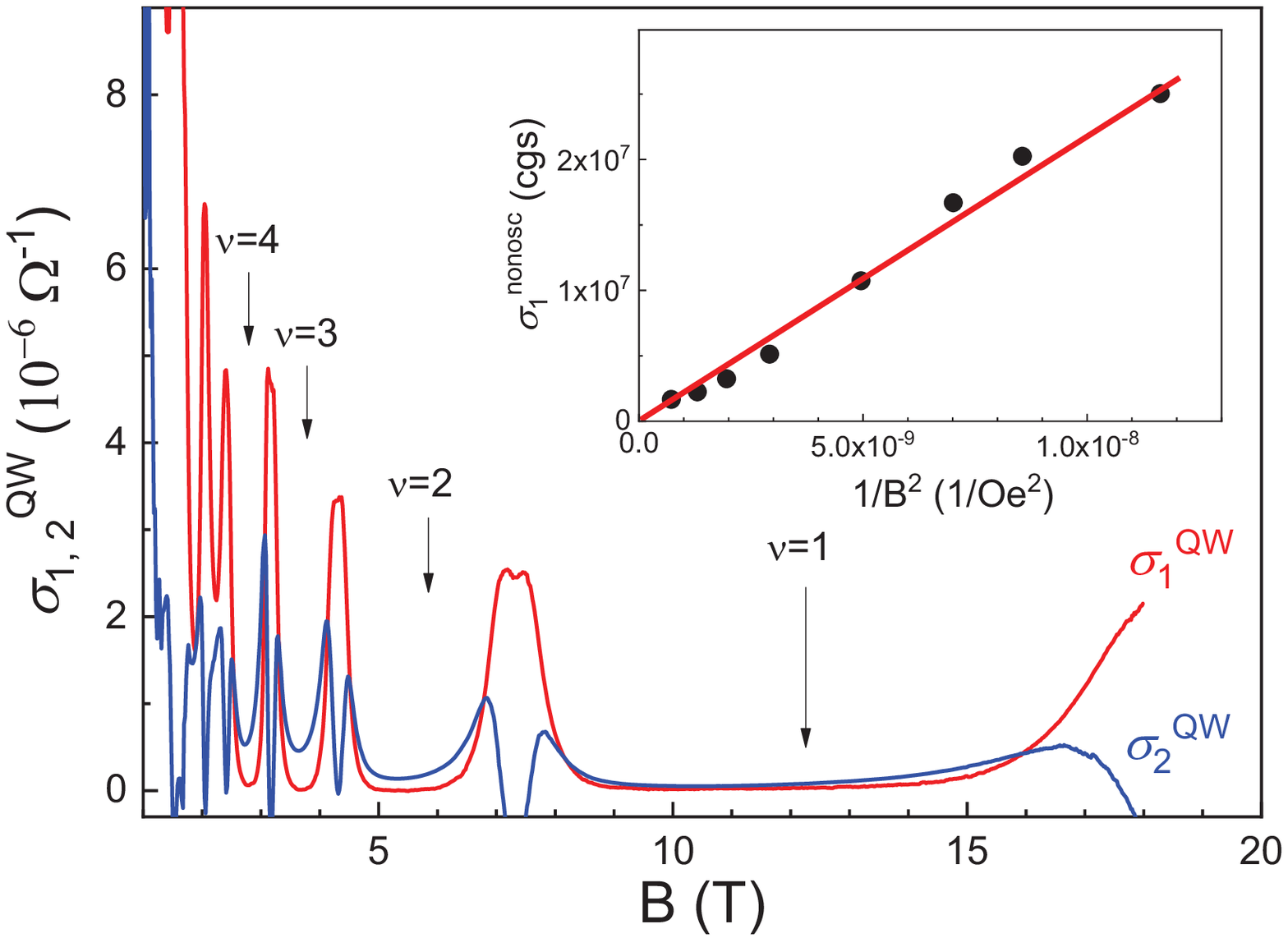}
\caption{Dependence of the conductance components $\sigma_1^{\text{QW}}$ and  $\sigma_2^{\text{QW}}$ in the quantum well on the magnetic field for frequency of 30 MHz, $T$=20~mK.
 }
\label{2DMobil}
\end{figure}

Now that $\Gamma (B)$ and $\Delta v/v (B)$ are known for the quantum well and the layer, one may calculate the real and imaginary components of the conductance in each of them. Shown in Fig.~\ref{2DMobil} are calculated magnetic field dependences of the real $\sigma_1^{\text{QW}}$ and imaginary $\sigma_2^{\text{QW}}$ components of the ac conductance in the quantum well.

As seen from the figure, the ac conductance in the quantum well  exhibits conventional Shubnikov-de Haas (SdH) oscillations in the low magnetic field region and transitions to the integer quantum Hall effect (IQHE) regime at high fields.

The ratio between the real and imaginary components of the ac conductance allows one to trace the evolution of the conduction mechanism in the quantum well with a magnetic field.
Indeed, at the maxima of the conductance oscillations, $\sigma_1^{\text{QW}} > \sigma_2^{\text{QW}}$, which is characteristic of conduction by delocalized electrons. In contrast, at the minima of the oscillations, $\sigma_2^{\text{QW}} > \sigma_1^{\text{QW}}$, indicating the  hopping conductivity of electrons localized in the minima of the random potential~\citep{Drichko2000}.

Additionally, we can calculate the electron mobility in the quantum well using the method described in~\citep{Drichko1997}: an average (non-oscillating) part of the ac conductance $\sigma_1^{nonosc}$ depends on the magnetic field at relatively low $B$ as
\begin{equation}
\sigma_1^{nonosc} (B) = \frac{\sigma_0}{1+(\mu B/c)^2}
\label{eq:mobil}
\end{equation}
where $\sigma_0$ is the conductivity of the quantum well at $B=0$, $\mu$ is the mobility, and $c = 3 \times 10^{10}$ cm/s.~\citep{Ando1974}

Therefore, we build the dependence of $\sigma_1^{nonosc}$ versus $1/B^2$  in the magnetic field $B < $5~T, see inset of Fig.~\ref{2DMobil}.
Provided that $(\mu B /c)^2 \gg$~1, the electron mobility in the quantum well can be determined from the slope of this linear dependence since the electron concentration $n$ is known from the position of the Shubnikov-de Haas oscillations in the magnetic field.

The mobility determined experimentally in this way was $\mu = 1.5 \times 10^5$ cm$^2$/Vs, which is close to the value of $\mu = 2 \times 10^5$ cm$^2$/Vs obtained from direct current measurements. This agreement confirms that the method of separating the attenuation and velocity contributions from the quantum well and the parallel conducting layer is justified.

\begin{figure}[h]
\centering
\includegraphics[width=0.9\linewidth]{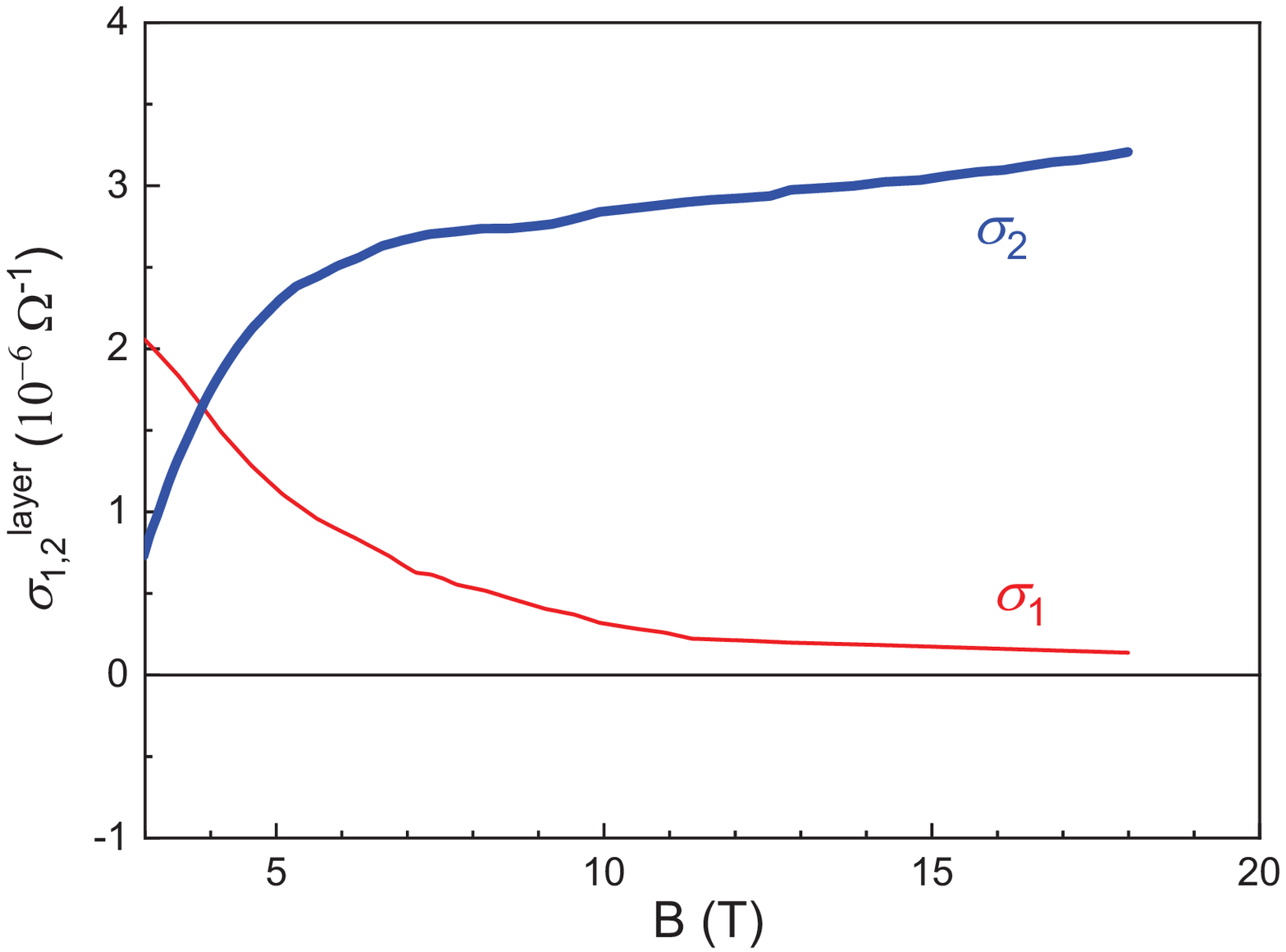}
\caption{Dependences of the ac conductance components $\sigma_1^{\text{layer}}$ and  $\sigma_2^{\text{layer}}$ in the shunting layer on the magnetic field. $T$=20~mK, $f$=85~MHz.
 }
\label{S12Layer}
\end{figure}

We can now calculate and analyze the ac conductivity in the shunting layer, see Fig.~\ref{S12Layer}.

It is evident from Fig.~\ref{S12Layer} that the conductivity component
$\sigma_2^{\text{layer}}$ increases with magnetic field and for
$B > 4$ T becomes larger than
$\sigma_1$. This further confirms the hopping nature of conduction in the layer. Simultaneously,
$\sigma_1^{\text{layer}}$, which is associated with delocalized electrons, decreases with increasing magnetic field. This decrease is likely caused by a reduction in electron density due to freeze-out and a decrease in electron mobility.

It should be noted that the decrease in conductivity with magnetic field increase could also be attributed to weak localization effects. However, in the studied structure, these effects are negligible~\citep{Zhi2025,Lehner2018}, having been observed only at magnetic fields below 1 T. This work discusses results obtained in fields above 2~T, where such effects are no longer significant.

As for hopping conduction, such a mechanism seems to be well described by a two-site model~\citep{Pollak1961}, where electron hopping occurs between states with close energies that are localized at two different impurity centers separated by a distance less than the average.

At the same time it should be noted that the described above  conductance separation procedure could not be performed for high frequencies,  $f >$~85~MHz. This suggests the possible existence of an additional mechanism that alters the magnetic field dependences of both sound absorption and velocity shift in the sample with the shunting layer. Since this effect is more pronounced at higher frequencies, it is reasonable to hypothesize that it may originate from, for example, sound scattering on certain inhomogeneities in the layer with characteristic sizes comparable to the acoustic wavelength (10$\div$20~$\mu$m).

\section{Measurements in a tilted magnetic field}

Despite the presence of the layer shunting the conduction through the quantum well, we can still determine the carriers $g$-factor using the so-called coincidence method~\citep{Brosig2000,Chokomakoua2004,Nedniyom2009} without singling out conductance in the quantum well.  That can be done safely because the hopping conductivity in the shunting layer does not exhibit Shubnikov-de Haas (SdH) oscillations.
To implement the coincidence method, we measured the SAW attenuation ($\Gamma$) and the relative velocity shift ($\Delta v/v$) in tilted magnetic fields up to 18~T at a SAW frequency of 85~MHz and a temperature of 0.3~K. A sample was mounted on a one-axis rotator, which enabled us to change the angle between the normal to the structure and the magnetic field. The tilt angle $\Theta$, defined relative to the normal to the sample surface, was varied from 0$^o$ to 90$^o$.

\begin{figure*}[t]
\centering
\includegraphics[width=0.8\linewidth]{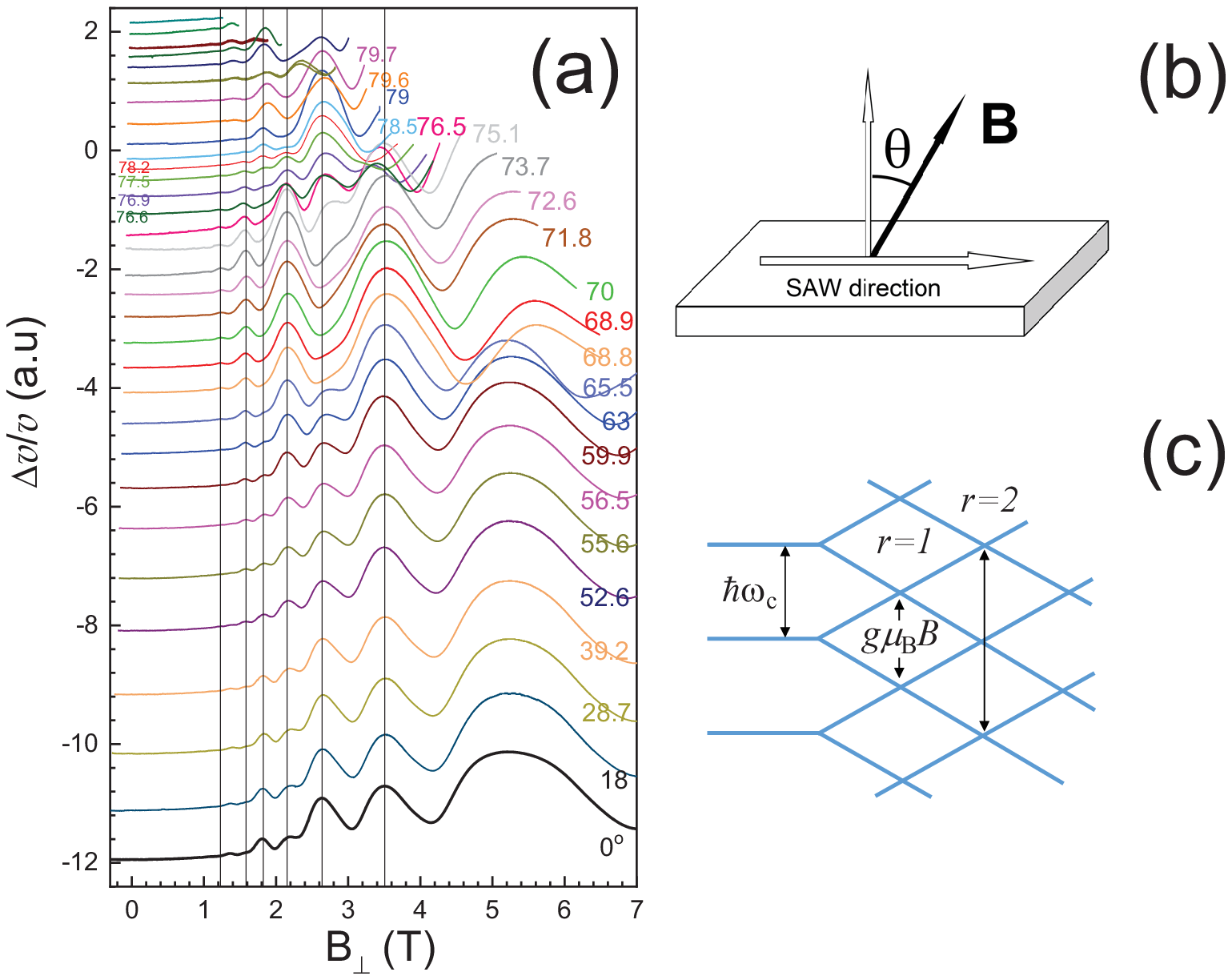}
\caption{a) Dependence of the SAW velocity on the perpendicular component $B_\perp$ of the total magnetic
field $B$ at varied tilt angles. Traces are offset for clarity. $T$=300~mK$; f$=85~MHz.
b) Tilt configuration.
c) Schematic of the energy levels.}
\label{fig7}
\end{figure*}

The coincidence method is based on the different field dependences of cyclotron and spin splitting energies. While the cyclotron frequency $\omega_c=eB/m^*c$ is determined by the component of the magnetic field perpendicular to the quantum well, the spin splitting $g \mu_B B_{tot}$ depends on the total field magnitude. This difference leads to Landau level crossings when the magnetic field is tilted at specific angles relative to the sample normal.

The angles at which crossings of spin-split Landau levels were observed manifested identically in both the $\Gamma$ and $\Delta v/v$ dependences on the perpendicular magnetic field component $B_{\bot}$. The velocity dependences on the magnetic field $B_{\bot}$ at different angles are presented in Fig.~\ref{fig7}.

Different orders of level coincidence are typically characterized by the parameter $r$, defined as the ratio of Zeeman to cyclotron energies:

\begin{equation}
\label{eq:r}
r = \frac{g\mu_BB_{tot}}{\hbar\omega_c} = \frac{gm^*}{2\cos\Theta}
\end{equation}
where $\mu_B=e\hbar/2m_0 c$ is the Bohr magneton, $\omega_c = eB/m^*c$ is the cyclotron frequency, $\Theta$ is the angle between the direction of the magnetic field and the surface normal.

For even $\nu$=4 and 6 the levels' crossings occur at $r$=1, while Landau levels cross at odd  $\nu$=3, 5, and 7  correspond to  $r$=2.  Having found $r$ and  $\Theta$, one can find the $g$-factor from Eq.~\ref{eq:r} if  the value of the effective mass is known.
However, since the energy band of $n$-InSb is non-parabolic, the value of the effective mass is usually determined individually for each sample  from the temperature dependence of the Shubnikov-de Haas oscillations. The latter is extracted from the Lifshitz-Kosevich formula~\citep{Lifshits1956}. Unfortunately, we cannot use this procedure due to low accuracy while in isolating the quantum well contribution to the SAW attenuation. Therefore, we used experimental data ($m=m^*\times m_0$) $m^*$=0.017 obtained on a similar sample~\citep{Lei2020}.

Since the $g$-factor turned out to depend on both
$r$ and the filling factor $\nu$, we attribute this dependence to its sensitivity to the spin polarization, which at the coincidence conditions is given by the following equation~\citep{Nedniyom2009}
\begin{equation}
\label{eq:P}
P = \frac{n\uparrow-n\downarrow}{n\uparrow+n\downarrow}=\frac{r}{\nu}.
\end{equation}
Where $n\uparrow$ and $n\downarrow$ are the densities of spin-up and spin-down electrons.

\begin{figure}[t]
\centering
\includegraphics[width=0.9\linewidth]{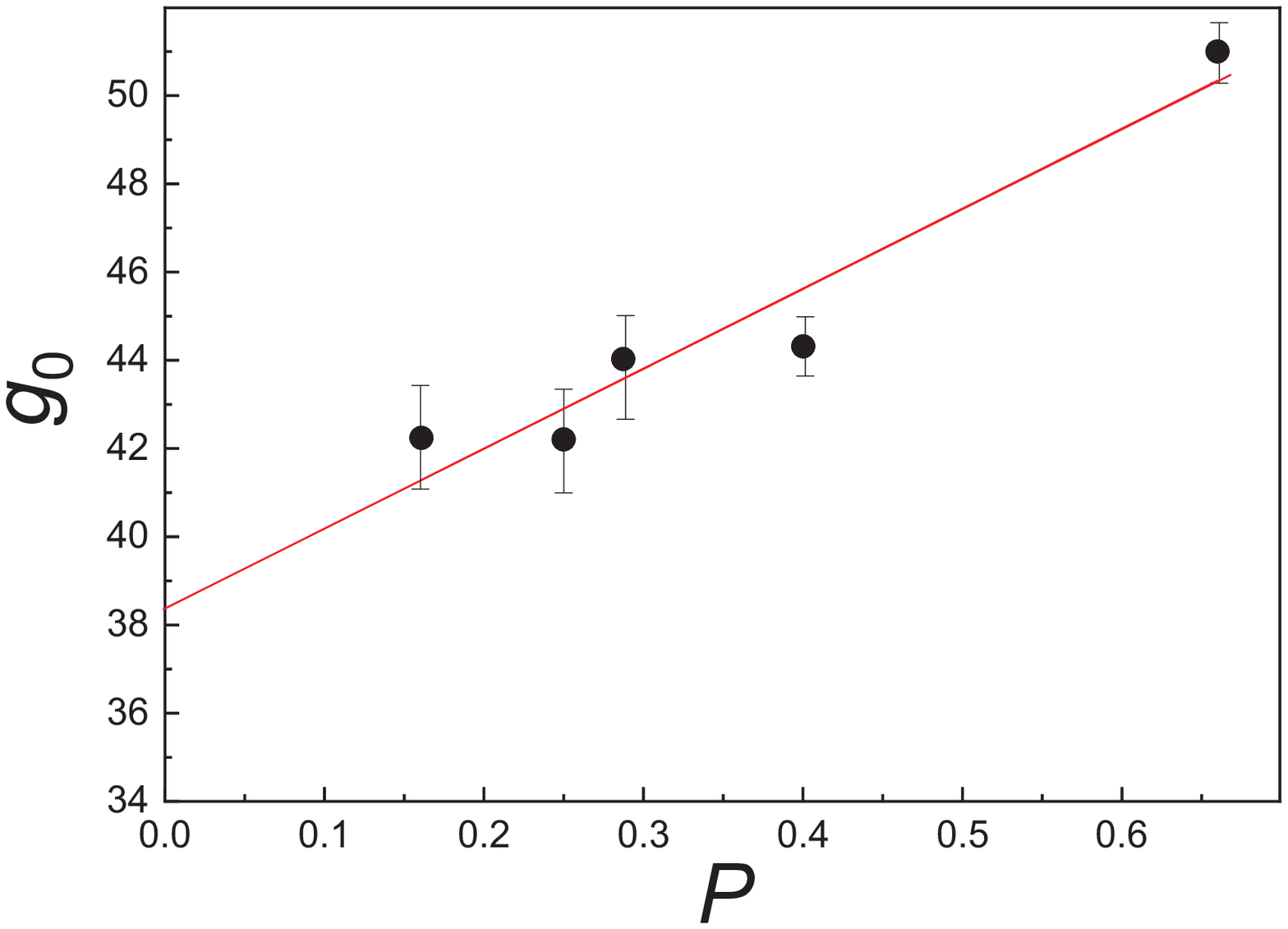}
\caption{The dependence of $g$-factor on the polarization $P$ }
\label{fig8}
\end{figure}

The dependence of the $g$-factor on spin polarization $P$ is shown in Figure~\ref{fig8}.

As seen in Figure~\ref{fig8}, the $g$-factor at zero polarization ($P$=0) is $g_0$=38. This value agrees well with $g_0$=35$\pm$4 reported in Ref.~\citep{Lei2020} for unpolarized conditions. Indeed, the authors of Ref.~\citep{Lei2020} determined $g_0$ from coincidence measurements at filling factors $\nu$=5$\div$28, corresponding to very weak polarization. In our current study, measurements were performed at $\nu$=3-7 where spin polarization becomes non-negligible.

The electron g-factor can therefore be expressed as having two components: $g_0$($P$=0), representing the polarization-independent constant term, and $g^*$ describing its polarization-dependent variation~\citep{Nedniyom2009,Yang2011}:
\begin{equation}
\label{eq:g}
g = g_0 + g^* \cdot P
\end{equation}

For the 21~nm wide $n$-InSb quantum well with carrier density $n=2.6\times 10^{11}$cm$^{-2}$ studied here, we determined $g_0=38\pm2$ and $g^*=20\pm2$.

\section{Conclusion} \label{Conclusion}

In $n$-InSb quantum well structures with a parallel conductive shunting layer, acoustic techniques enable separation of conductance contributions from the quantum well and the shunting layer.

It was found that the conductance $\sigma_1$ in the quantum well oscillates in a magnetic field. In low magnetic fields, these are Shubnikov-de Haas oscillations, and in high fields the integer quantum Hall effect regime is observed.


In the quantum well at the minima of the conductivity oscillations, electrons become localized in the random potential. Under these conditions, conduction occurs via electron hopping between these localized states and $\sigma_2^{\text{QW}} > \sigma_1^{\text{QW}}$.

While in the quantum well the transition of electrons from delocalized to localized states occurs with a change in the magnetic field, in the shunting layer the majority of electrons remain localized in magnetic fields $B >$4~T. Consequently, charge transport in the shunting layer in this magnetic field occurs via hopping between localized states, so $\sigma_2^{\text{layer}} > \sigma_1^{\text{layer}}$.

The coincidence method was employed to determine the electron g-factor in the quantum well and to investigate its dependence on spin polarization. It was found that the presence of a conducting layer does not affect the determination of the g-factor.

\section{Acknowledgements} \label{Acknowledgements}

The National High Magnetic Field Laboratory is supported by the National Science Foundation through NSF/DMR-2128556 and the State of Florida. We are thankful to R. Nowell for help with the experiment. The authors are grateful to Christian A. Lehner  for provided InSb QW samples.


\begin{thebibliography}{19}
\expandafter\ifx\csname natexlab\endcsname\relax\def\natexlab#1{#1}\fi
\providecommand{\url}[1]{\texttt{#1}}
\providecommand{\href}[2]{#2}
\providecommand{\path}[1]{#1}
\providecommand{\DOIprefix}{doi:}
\providecommand{\ArXivprefix}{arXiv:}
\providecommand{\URLprefix}{URL: }
\providecommand{\Pubmedprefix}{pmid:}
\providecommand{\doi}[1]{\href{http://dx.doi.org/#1}{\path{#1}}}
\providecommand{\Pubmed}[1]{\href{pmid:#1}{\path{#1}}}
\providecommand{\bibinfo}[2]{#2}
\ifx\xfnm\relax \def\xfnm[#1]{\unskip,\space#1}\fi
\bibitem[{Ando(1974)}]{Ando1974}
\bibinfo{author}{Ando, T.}, \bibinfo{year}{1974}.
\newblock \bibinfo{title}{Theory of quantum transport in a two-dimensional
  electron system under magnetic fields. {IV}. oscillatory conductivity}.
\newblock \bibinfo{journal}{J. Phys. Soc. Jpn.} \bibinfo{volume}{37},
  \bibinfo{pages}{1233--1239}.
\bibitem[{Brosig et~al.(2000)Brosig, Ensslin, Jansen, Nguyen, Brar, Thomas and
  Kroemer}]{Brosig2000}
\bibinfo{author}{Brosig, S.}, \bibinfo{author}{Ensslin, K.},
  \bibinfo{author}{Jansen, A.G.}, \bibinfo{author}{Nguyen, C.},
  \bibinfo{author}{Brar, B.}, \bibinfo{author}{Thomas, M.},
  \bibinfo{author}{Kroemer, H.}, \bibinfo{year}{2000}.
\newblock \bibinfo{title}{{InAs-AlSb} quantum wells in tilted magnetic fields}.
\newblock \bibinfo{journal}{Phys. Rev. B} \bibinfo{volume}{61},
  \bibinfo{pages}{13045--13048}.
\newblock \DOIprefix\doi{10.1103/PhysRevB.61.13045}.
\bibitem[{Chokomakoua et~al.(2004)Chokomakoua, Goel, Chung, Santos, Hicks,
  Johnson and Murphy}]{Chokomakoua2004}
\bibinfo{author}{Chokomakoua, J.C.}, \bibinfo{author}{Goel, N.},
  \bibinfo{author}{Chung, S.J.}, \bibinfo{author}{Santos, M.B.},
  \bibinfo{author}{Hicks, J.L.}, \bibinfo{author}{Johnson, M.B.},
  \bibinfo{author}{Murphy, S.Q.}, \bibinfo{year}{2004}.
\newblock \bibinfo{title}{Ising quantum hall ferromagnetism in {InSb}-based
  two-dimensional electronic systems}.
\newblock \bibinfo{journal}{Phys. Rev. B} \bibinfo{volume}{69},
  \bibinfo{pages}{235315}.
\newblock \DOIprefix\doi{10.1103/PhysRevB.69.235315}.
\bibitem[{Drichko et~al.(2005a)Drichko, Diakonov, Smirnov, Andrianov, Mironov,
  Myronov, Leadley and Whall}]{Drichko2005}
\bibinfo{author}{Drichko, I.L.}, \bibinfo{author}{Diakonov, A.M.},
  \bibinfo{author}{Smirnov, I.Y.}, \bibinfo{author}{Andrianov, G.O.},
  \bibinfo{author}{Mironov, O.A.}, \bibinfo{author}{Myronov, M.},
  \bibinfo{author}{Leadley, D.R.}, \bibinfo{author}{Whall, T.E.},
  \bibinfo{year}{2005}a.
\newblock \bibinfo{title}{High-frequency transport in p-type {Si/SiGe}
  heterostructures studied with surface acoustic waves in the quantum hall
  regime}.
\newblock \bibinfo{journal}{Phys. Rev. B} \bibinfo{volume}{71},
  \bibinfo{pages}{045333}.
\newblock \DOIprefix\doi{10.1103/PhysRevB.71.045333}.
\bibitem[{Drichko et~al.(2000)Drichko, Diakonov, Smirnov, Galperin and
  Toropov}]{Drichko2000}
\bibinfo{author}{Drichko, I.L.}, \bibinfo{author}{Diakonov, A.M.},
  \bibinfo{author}{Smirnov, I.Y.}, \bibinfo{author}{Galperin, Y.M.},
  \bibinfo{author}{Toropov, A.I.}, \bibinfo{year}{2000}.
\newblock \bibinfo{title}{High-frequency hopping conductivity in the quantum
  hall effect regime: {Acoustical} studies}.
\newblock \bibinfo{journal}{Phys. Rev. B} \bibinfo{volume}{62},
  \bibinfo{pages}{7470--7477}.
\newblock \DOIprefix\doi{10.1103/PhysRevB.62.7470}.
\bibitem[{Drichko et~al.(2005b)Drichko, Dyakonov, Smirnov, Suslov, Galperin,
  Yakimov and Nikiforov}]{Drichko2005a}
\bibinfo{author}{Drichko, I.L.}, \bibinfo{author}{Dyakonov, A.M.},
  \bibinfo{author}{Smirnov, I.Y.}, \bibinfo{author}{Suslov, A.V.},
  \bibinfo{author}{Galperin, Y.M.}, \bibinfo{author}{Yakimov, A.I.},
  \bibinfo{author}{Nikiforov, A.I.}, \bibinfo{year}{2005}b.
\newblock \bibinfo{title}{Mechanisms of low-temperature high-frequency
  conductivity in systems with a dense array of {Ge$_{0.7}$Si$_{0.3}$} quantum
  dots in silicon}.
\newblock \bibinfo{journal}{J. Exp. Theor. Phys.} \bibinfo{volume}{101},
  \bibinfo{pages}{1122--1132}.
\newblock \DOIprefix\doi{10.1134/1.2163927}.
\bibitem[{Drichko et~al.(2013)Drichko, Malysh, Smirnov, Suslov, Mironov, Kummer
  and von Känel}]{Drichko2013}
\bibinfo{author}{Drichko, I.L.}, \bibinfo{author}{Malysh, V.A.},
  \bibinfo{author}{Smirnov, I.Y.}, \bibinfo{author}{Suslov, A.V.},
  \bibinfo{author}{Mironov, O.A.}, \bibinfo{author}{Kummer, M.},
  \bibinfo{author}{von Känel, H.}, \bibinfo{year}{2013}.
\newblock \bibinfo{title}{Acoustoelectric effects in very high-mobility
  p-{SiGe/Ge/SiGe} heterostructure at low temperatures in high magnetic
  fields}.
\newblock \bibinfo{journal}{J. Appl. Phys.} \bibinfo{volume}{114},
  \bibinfo{pages}{074302}.
\newblock \DOIprefix\doi{10.1063/1.4818436}.
\bibitem[{Drichko and Smirnov(1997)}]{Drichko1997}
\bibinfo{author}{Drichko, I.L.}, \bibinfo{author}{Smirnov, I.Y.},
  \bibinfo{year}{1997}.
\newblock \bibinfo{title}{Contact-free determination of the parameters of a 2d
  electron gas in {GaAs/AlGaAs} heterostructures}.
\newblock \bibinfo{journal}{Semiconductors} \bibinfo{volume}{31},
  \bibinfo{pages}{933--936}.
\bibitem[{Drichko et~al.(2011)Drichko, Smirnov, Suslov and
  Leadley}]{Drichko2011}
\bibinfo{author}{Drichko, I.L.}, \bibinfo{author}{Smirnov, I.Y.},
  \bibinfo{author}{Suslov, A.V.}, \bibinfo{author}{Leadley, D.R.},
  \bibinfo{year}{2011}.
\newblock \bibinfo{title}{Acoustic studies of ac conductivity mechanisms in
  {GaAs/AlGaAs} in the integer and fractional quantum hall effect regime}.
\newblock \bibinfo{journal}{Phys. Rev. B} \bibinfo{volume}{83},
  \bibinfo{pages}{235318}.
\newblock \DOIprefix\doi{10.1103/PhysRevB.83.235318}.
\bibitem[{Galperin and Priev(1986)}]{Galperin1986}
\bibinfo{author}{Galperin, Y.M.}, \bibinfo{author}{Priev, E.Y.},
  \bibinfo{year}{1986}.
\newblock \bibinfo{title}{Acoustical properties of semiconductors in hopping
  conductivity regime}.
\newblock \bibinfo{journal}{Sov. Phys. Solid State} \bibinfo{volume}{28},
  \bibinfo{pages}{385--389}.
\bibitem[{Kagan(1997)}]{Kagan1997}
\bibinfo{author}{Kagan, V.D.}, \bibinfo{year}{1997}.
\newblock \bibinfo{title}{Propagation of a surface acoustic wave in a layered
  system containing a two-dimensional conducting layer}.
\newblock \bibinfo{journal}{Semiconductors} \bibinfo{volume}{31},
  \bibinfo{pages}{407--411}.
\newblock \DOIprefix\doi{10.1134/1.1187321}.
\bibitem[{Khodaparast et~al.(2004)Khodaparast, Doezema, Chung, Goldammer and
  Santos}]{Khodaparast2004}
\bibinfo{author}{Khodaparast, G.A.}, \bibinfo{author}{Doezema, R.E.},
  \bibinfo{author}{Chung, S.J.}, \bibinfo{author}{Goldammer, K.J.},
  \bibinfo{author}{Santos, M.B.}, \bibinfo{year}{2004}.
\newblock \bibinfo{title}{Spectroscopy of rashba spin splitting in insb quantum
  wells}.
\newblock \bibinfo{journal}{Phys. Rev. B} \bibinfo{volume}{70},
  \bibinfo{pages}{155322}.
\newblock \DOIprefix\doi{10.1103/PhysRevB.70.155322}.
\bibitem[{Lehner et~al.(2018)Lehner, Tschirky, Ihn, Dietsche, Keller, Fält and
  Wegscheider}]{Lehner2018}
\bibinfo{author}{Lehner, C.A.}, \bibinfo{author}{Tschirky, T.},
  \bibinfo{author}{Ihn, T.}, \bibinfo{author}{Dietsche, W.},
  \bibinfo{author}{Keller, J.}, \bibinfo{author}{Fält, S.},
  \bibinfo{author}{Wegscheider, W.}, \bibinfo{year}{2018}.
\newblock \bibinfo{title}{Limiting scattering processes in high-mobility {InSb}
  quantum wells grown on {GaSb} buffer systems}.
\newblock \bibinfo{journal}{Phys. Rev. Mater.} \bibinfo{volume}{2},
  \bibinfo{pages}{054601}.
\newblock \DOIprefix\doi{10.1103/PhysRevMaterials.2.054601}.
\bibitem[{Lei et~al.(2020)Lei, Lehner, Rubi, Cheah, Karalic, Mittag, Alt,
  Scharnetzky, Märki, Zeitler, Wegscheider, Ihn and Ensslin}]{Lei2020}
\bibinfo{author}{Lei, Z.}, \bibinfo{author}{Lehner, C.A.},
  \bibinfo{author}{Rubi, K.}, \bibinfo{author}{Cheah, E.},
  \bibinfo{author}{Karalic, M.}, \bibinfo{author}{Mittag, C.},
  \bibinfo{author}{Alt, L.}, \bibinfo{author}{Scharnetzky, J.},
  \bibinfo{author}{Märki, P.}, \bibinfo{author}{Zeitler, U.},
  \bibinfo{author}{Wegscheider, W.}, \bibinfo{author}{Ihn, T.},
  \bibinfo{author}{Ensslin, K.}, \bibinfo{year}{2020}.
\newblock \bibinfo{title}{Electronic g factor and magnetotransport in {InSb}
  quantum wells}.
\newblock \bibinfo{journal}{Phys. Rev. Res.} \bibinfo{volume}{2},
  \bibinfo{pages}{033213}.
\newblock \DOIprefix\doi{10.1103/PhysRevResearch.2.033213}.
\bibitem[{Lifshits and Kosevich(1956)}]{Lifshits1956}
\bibinfo{author}{Lifshits, I.M.}, \bibinfo{author}{Kosevich, A.M.},
  \bibinfo{year}{1956}.
\newblock \bibinfo{title}{Theory of magnetic susceptibility in metals at low
  temperature}.
\newblock \bibinfo{journal}{Sov. Phys. JETP} \bibinfo{volume}{2},
  \bibinfo{pages}{636--645}.
\bibitem[{Nedniyom et~al.(2009)Nedniyom, Nicholas, Emeny, Buckle, Gilbertson,
  Buckle and Ashley}]{Nedniyom2009}
\bibinfo{author}{Nedniyom, B.}, \bibinfo{author}{Nicholas, R.J.},
  \bibinfo{author}{Emeny, M.T.}, \bibinfo{author}{Buckle, L.},
  \bibinfo{author}{Gilbertson, A.M.}, \bibinfo{author}{Buckle, P.D.},
  \bibinfo{author}{Ashley, T.}, \bibinfo{year}{2009}.
\newblock \bibinfo{title}{Giant enhanced g-factors in an {InSb} two-dimensional
  gas}.
\newblock \bibinfo{journal}{Phys. Rev. B} \bibinfo{volume}{80},
  \bibinfo{pages}{125328}.
\newblock \DOIprefix\doi{10.1103/PhysRevB.80.125328}.
\bibitem[{Pollak and Geballe(1961)}]{Pollak1961}
\bibinfo{author}{Pollak, M.}, \bibinfo{author}{Geballe, T.H.},
  \bibinfo{year}{1961}.
\newblock \bibinfo{title}{Low-frequency conductivity due to hopping processes
  in silicon}.
\newblock \bibinfo{journal}{Phys. Rev.} \bibinfo{volume}{122},
  \bibinfo{pages}{1742--1746}.
\newblock \DOIprefix\doi{10.1103/PhysRev.122.1742}.
\bibitem[{Yang et~al.(2011)Yang, Liu, Mishima, Santos, Nagase and
  Hirayama}]{Yang2011}
\bibinfo{author}{Yang, K.F.}, \bibinfo{author}{Liu, H.W.},
  \bibinfo{author}{Mishima, T.D.}, \bibinfo{author}{Santos, M.B.},
  \bibinfo{author}{Nagase, K.}, \bibinfo{author}{Hirayama, Y.},
  \bibinfo{year}{2011}.
\newblock \bibinfo{title}{Nonlinear magnetic field dependence of spin
  polarization in high-density two-dimensional electron systems}.
\newblock \bibinfo{journal}{New J. Phys.} \bibinfo{volume}{13},
  \bibinfo{pages}{083010}.
\newblock \DOIprefix\doi{10.1088/1367-2630/13/8/083010}.
\bibitem[{Zhi et~al.(2025)Zhi, Wu, Ruan, Liu, Huang, Yao, Liu, Tang, Yao, Sun,
  Zhang, Xiao, Che and Kou}]{Zhi2025}
\bibinfo{author}{Zhi, Z.}, \bibinfo{author}{Wu, Y.}, \bibinfo{author}{Ruan,
  H.}, \bibinfo{author}{Liu, J.}, \bibinfo{author}{Huang, P.},
  \bibinfo{author}{Yao, S.}, \bibinfo{author}{Liu, X.}, \bibinfo{author}{Tang,
  C.}, \bibinfo{author}{Yao, Q.}, \bibinfo{author}{Sun, L.},
  \bibinfo{author}{Zhang, Y.}, \bibinfo{author}{Xiao, Y.},
  \bibinfo{author}{Che, R.}, \bibinfo{author}{Kou, X.}, \bibinfo{year}{2025}.
\newblock \bibinfo{title}{Tunable interfacial rashba spin–orbit coupling in
  asymmetric {AlInSb/InSb/CdTe} quantum well heterostructures}.
\newblock \bibinfo{journal}{Appl. Phys. Lett.} \bibinfo{volume}{126},
  \bibinfo{pages}{012104}.
\newblock \DOIprefix\doi{10.1063/5.0233964}.

\end{thebibliography}
\end{document}